\documentstyle[amsmath,amssymb,aps,eqsecnum,graphicx]{revtex}

\begin{document}
\draft
\title{Manifolds of interconvertible pure states
}
\author{ Magdalena M. Sino{\l}\c{e}cka, Karol {\.Z}yczkowski\footnote{also at 
 Instytut Fizyki,  Uniwersytet Jagiello{\'n}ski, ul. Reymonta 4,
30-059 Krak{\'o}w, Poland}, Marek Ku\'s } 
\address{
Center for Theoretical Physics, Polish Academy of Sciences,
\\ Al. Lotnik{\'o}w 32/44, 02-668 Warszawa, Poland}
\date{October 12, 2001}
\maketitle

\begin{abstract}
Local orbits of a pure state of an $N \times N$  bi-partite quantum system are analyzed. We 
compute their dimensions which depends on the degeneracy of the vector of coefficients arising 
by the Schmidt decomposition. In particular, the generic orbit has $2N^2 -N-1$ dimensions, the 
set of separable states is $4(N-1)$ dimensional, while the manifold of maximally entangled 
states has $N^2-1$ dimensions.  \\[0.2cm]%
\end{abstract}

\section{Introduction}

The existence of entangled states, i.e., roughly speaking, the states of a
composite system which exhibit quantum correlations among the subsystems,
appeared recently to be extremely important in rapidly developing field of
quantum communication. It is due to non-classical properties of entangled
states that various schemes of quantum computing, quantum cryptography and
quantum teleportation can be thought of being practically realizable.

A pure state $|\psi\rangle$ in the Hilbert space ${\cal H}={\cal H}_A
\otimes
{\cal H}_B$ of a composite quantum system consisting of two subsystems $A$
and
$B$ with Hilbert spaces ${\cal H}_A$ and ${\cal H}_B$ is separable, if it
can
be cast to the product form
$|\psi\rangle=|\psi_A\rangle\otimes|\psi_B\rangle$,
where $|\psi_A\rangle$ and $|\psi_B\rangle$ are some states of the
subsystems.
States which are not separable are called {\sl entangled}. The situation is
more complicated in the case of a mixed state (a density matrix $\rho$)
\cite{werner}. It is separable if it is expressible as a convex sum of
product
states: $\rho=\sum_ip_i\rho_i^{(A)}\otimes\rho_i^{(B)}$, $p_i>0$,
$\sum_ip_i=1$, where $\rho_i^{(A)}$ and $\rho_i^{(B)}$ are, in general
mixed,
states of the subsystems. A mixed state is called entangled if it is not
separable. In what follows we consider only systems with finite-dimensional
Hilbert spaces which seem to be more important in proposed applications of
quantum information theory, the infinite-dimensional case needs some
refinement
of the above definition of separability.

It is relatively easy to check whether a given pure state is separable or
entangled (e.g.\ by investigating its Schmidt coefficients - see below). The
situation complicates for mixed states - we do not know how to check
unambiguously separability of a given mixed states if the dimensionality of
the Hilbert spaces of subsystems exceeds $3$ \cite{primer}.

As a problem complementary to determining the separability properties of a given state one can 
pose the question of the relation between the set of the separable (entangled) states to the set 
of all states of the composite system. This can be understood as the question of a relative 
measure of the set of entangled states (i.e. "how probable is that a given state is entangled?") 
- the problem posed and partially  solved in \cite{zyckowski:98,Zy99}, or about the geometrical 
and topological properties of this set. In this paper we concentrate on the latter problem in 
the following setting. Since we are interested in quantum correlations between two subsystems we 
should take into consideration only these properties which do not change under various quantum 
mechanical operations performed locally in each subsystem. Thus two states which are 
interconvertible one to another via local unitary transformations (i.e. purely quantum 
mechanical operations without decoherence) are equivalent from the point of their entanglement 
properties. This can lead to construction of appropriate measures of entanglement characterizing 
the classes of equivalent states. Our approach is in a sense complementary to the task of 
identifying the set of all {\sl invariants} with respect to the local unitary transformations 
\cite{LPS99,GRB98,EM99,CS00,Ma00,Lo01,AF01}. 

In this work we pose and solve the question of the dimensionality and
topology of manifolds of
states equivalent to a given one via local unitary transformations. Thus the
present paper may
be regarded as an extension of \cite{KZ01} (see also \cite{BH01,BBZ01,MD01}),
in which these
questions were discussed for the simplest system of two qubits. In the case
of pure states we
find the explicit results for any $ N \times N $ composite system by
identifying explicitly the
topology of the orbits as well as in a purely algebraic, algorithmic manner.
The second approach
which does not depend on the Schmidt decomposition (see below) is, in
principle, applicable also
to mixed states, this is illustrated by considering the generalized Werner
states \cite{werner}.

\section{Pure entangled states}

\subsection{ Schmidt decomposition}

Consider a pure state $|\psi\rangle$ of a composite Hilbert space ${\cal
H}={\cal H}_A \otimes {\cal H}_B$ of size $N^2$. Introducing an orthonormal
basis $\{ |n\rangle \}_{n=1}^N$ in each subsystem, we may represent the
state
as
\begin{equation}
 |\psi\rangle = \sum_{n=1}^N
 \sum_{m=1}^N C_{mn} |n\rangle \otimes |m\rangle.
\label{state}
\end{equation} The complex matrix of coefficients $C$ of size $N$
needs not to be Hermitian nor normal. Its singular values (i.e. the square
roots of eigenvalues $\lambda_k$ of the positive matrix $C^{\dagger}C$)
determine the Schmidt decomposition \cite{Sc06,Pe93,EK95}
\begin{equation}
|\psi\rangle = \sum_{k=1}^{N} \sqrt{\lambda_{k}}
 |k^{\prime}\rangle
\otimes |k^{\prime\prime}\rangle,
 \label{VSchmidt}
\end{equation}
where the basis in $\cal H$  is transformed by a local unitary transformation $W \otimes V$. 
Thus $|k^{\prime}\rangle = W |k\rangle$, and $|k^{\prime\prime}\rangle = V |k\rangle$, where $W$ 
and $V$ are the matrices of eigenvectors of $C^{\dagger}C$ and $CC^{\dagger}$, respectively. In 
the generic case of a non-degenerate vector $\Lambda$, the Schmidt decomposition is unique up to 
two unitary diagonal matrices, up to which the matrices of eigenvectors $W$ and $V$ are 
determined. The normalization condition $\langle \psi|\psi\rangle=1$ enforces 
$\sum_{k=1}^{N}\lambda_k=1$. Thus the vector ${\Lambda}=(\lambda_1,...,\lambda_N)$ lives in the 
($N-1$) dimensional simplex ${\cal S}_N$. The Schmidt coefficients $\lambda_k$ do not depend on 
the initial basis $|n\rangle \otimes |m\rangle$, in which the analyzed state $|\psi\rangle$ is 
represented.

\subsection{ Pure state entanglement}

The Schmidt coefficient of a pure state  $|\psi\rangle$ are equal to the eigenvalues of the 
reduced density operator, obtained by partial tracing, $\rho^A={\rm tr}_B(|\psi\rangle \langle 
\psi|)$. A pure state is called {\sl separable}, if it can be represented in the product form 
$|\psi\rangle=|\psi_A\rangle \otimes |\psi_B\rangle$, where $|\psi_A\rangle \in {\cal H}_A$ and 
$|\psi_B\rangle\in {\cal H}_B$. This occurs if and only if there exists only one non-zero 
Schmidt coefficient, $\lambda_1=1$, i.e. the reduced state $\rho^{A}$ is pure. In the opposite 
case state $|\psi\rangle $ is called {\sl entangled}.    
 A pure state is called {\sl maximally entangled} if all its Schmidt coefficients are equal, 
 $\lambda_1=\lambda_k=1/N$. Note that the Schmidt coefficients are invariant with respect to any 
{\sl local operations} $U_L=U_A \otimes U_B$, and thus they may serve as ingredients of any 
measure of entanglement. 

\subsection{ Local orbits}

We are going to study the orbits of a given pure state $|\psi\rangle$ with
respect to the local transformations $U_L$. Two states belonging to the same
orbit are called {\sl interconvertible}, since they may be reversibly
transformed by local transformations one into another \cite{JP99}. Let us
order
its Schmidt coefficients $\Lambda=(0\leq\lambda_1\leq\lambda_2\leq\cdots
\leq\lambda_N)$. In order to describe the character of the degeneracy we
rename
them into $\Lambda=(0,\cdots,0,\nu_1,\cdots,\nu_1,\nu_2,\cdots,\nu_2,\dots,
\nu_K,\cdots,\nu_K)$ where each value $\nu_n$ occurs $m_n$ times and $m_0$
is
the number of vanishing Schmidt coefficients. Obviously $m_0+\sum_{n=1}^K
m_n=N$, and $m_0$ might be equal to zero. The main result of our paper is
contained in the following

{\sl Proposition. The local orbit generated from $|\psi\rangle$ has the
structure of the following quotient space}
\begin{equation}
{\cal O}=\frac{U(N)\times U(N)}{{\mathcal
G}(m_0,m_1,\ldots,m_K)},
\label{orbit}
\end{equation}
{\sl where ${\mathcal G}(m_0,m_1,\ldots,m_K)$ is the subgroup of the direct
product
$U(N)\times U(N)$ consisting of the pairs of unitary matrices $(U,V)$ of the
form}
\begin{equation}\label{u1v1}
U=\left[
\begin{array}{cccc}
  u_0 &     &        &  \\
      & u_1 &        &  \\
      &     &
\ddots & \\
      &     &        &  u_K
\end{array}
\right], \quad
V=e^{i\phi}\left[
\begin{array}{cccc}
  v_0 &       &        &  \\
      & u_1^* &        &  \\
      &       & \ddots & \\
      &       &        &  u_K^*
\end{array}
\right],
\end{equation}
{\sl where $u_0$ and $v_0$ arbitrary matrices from $U(m_0)$, and
$u_1,\ldots,u_K$ denote
arbitrary matrices from, respectively, $U(m_1),\ldots,U(m_K)$. The overall
phase factor
$e^{i\phi}$ accounts for the irrelevant phase of the state $|\psi\rangle$,
ie.\ we identify
states differing by a phase factor. The dimension of the orbit (\ref{orbit})
reads}
\begin{equation}
{\rm dim}({\cal O})= 2N^2-2m_0^2-\sum_{n=1}^K m_n^2-1 \ . \label{dimorb}
\end{equation}
Indeed, let us observe that the action of the tensor product  $U\otimes V
\in U(N)\otimes U(N)$
on the state (\ref{state}),
\begin{eqnarray}
U\otimes V |\psi\rangle=\sum_{m,n}C_{mn}U|m\rangle\otimes V|n\rangle
=\sum_{m,n,k,l}C_{mn}U_{km}|k\rangle\otimes V_{ln}|l\rangle
=\sum_{k,l}(UCV^T)_{kl}|k\rangle\otimes|l\rangle,
\end{eqnarray}
reduces to the direct product action on the coefficient matrix $C$
\begin{equation}
U(N)\times U(N) \ni (U,V):C\mapsto (U,V)(C):=UCV^T
\end{equation}
Let now the action of $(\tilde{U},\tilde{V})$ reduces $C$ to its diagonal
Schmidt form
\begin{equation}
\tilde{U}C\tilde{V}^T=
\text{diag}(0,\ldots,0,\nu_1,\ldots,\nu_2,\ldots,\nu_K,\ldots,\nu_K).
\end{equation}
Then $\tilde{U}C\tilde{V}^T=U\tilde{U}C\tilde{V}^TV^T$ iff $U$ and $V$ are
given by (\ref{u1v1})
Now the formula (\ref{orbit}) follows in an obvious manner, once we realize
that in fact we
should disregard any unimportant overall phase of (\ref{state}) (or in other
words we should
identify the coefficient matrices $C$ and $C'=e^{i\theta}C$, ie.\ work in an
appropriate
projective space). The dimension formula (\ref{dimorb}) follows from a
simple calculations
involving the dimensionalities of the unitary groups, while the last term equal
  to unity stems from the projectivisation procedure.
 An alternative, algebraic
proof of this result is proved in Section \ref{sec:Gram}.

In fact the orbit has a structure of a Cartesian product:
\begin{equation}\label{trivbundle}
{\cal O}=\frac {U(N)}{U(m_0)\times U(m_1)\times \cdots \times U(m_K)}
 \times \frac {U(N)}{U(m_0)\times U(1)},
\end{equation}
where the first factor represents global orbits in the set
      of density matrices of size N with the same spectrum
\cite{ACH93,ZS00}. In the language of fiber bundles such orbits form the base, while the fibers 
consists of all $N \times N$  pure states, which are related by partial tracing to a given 
density matrix of size $N$. We shall provide a complete proof of this fact elsewhere \cite{Kus}. 

In the generic case of all coefficients different (and non zero), i.e. $K=N$
the manifold is thus identified as
\begin{equation}
{\cal O}_g=\frac {U(N)}{[U(1)]^N}
 \times \frac {U(N)}{U(1)},
 \label{orbgen}
\end{equation}
with the dimension
\begin{equation}
{\rm dim}({\cal O}_g) = 2N^2-N-1  \ .
 \label{dimgen}
\end{equation}

The set of all  orbits enumerated above produces the complex projective
space
${\mathbb C}P^{N^2-1}$ -  the $(2N^2-2)$ dimensional manifold of pure states
of
the $N \times N$ system. However, the the set constructed of the generic
orbits
(\ref{dimgen}) generated by each point of the interior of the Weyl chamber,
is
of full measure in the space of pure states. In this way we demonstrated a
foliation of ${\mathbb C}P^{N^2-1}$. This foliation is {\sl singular}, since
there exist also (measure zero) leaves of various dimensions and topology,
as listed in Table 1 for $N=2,3$ and $4$.

\subsection{ Special cases: separable and maximally entangled states}

For separable states there exists only one non zero coefficient,
$\lambda_1=1$,
so $m_0=N-1$. Thus (\ref{orbit}) gives
\begin{equation}
 {\cal O}_{\rm sep}=\frac {U(N)}{U(1)\times U(N-1)} \times
  \frac{U(N)}{U(1)\times U(N-1)}=
  {\mathbb C}P^{N-1} \times {\mathbb C}P^{N-1},
 \label{orbsep}
\end{equation}
with the dimension ${\rm dim}({\cal O}_{\rm sep}) = 4(N-1)$. The maximally
entangled states are characterized by $\lambda_1=\lambda_N=1/N$, hence
$m_1=N$
and $m_0=0$. Therefore
\begin{equation}
 {\cal O}_{\rm max}=\frac{U(N)}{U(1)}=\frac{SU(N)}{Z_N},
 \label{orbmax}
\end{equation}
with the dimension ${\rm dim}({\cal O}_{max}) = N^2-1$. Note that this space is not isomorphic 
with $SU(N)$ because $U(N)$ is not a direct product of $U(1)$ and $SU(N)$ \cite{Mi80}. Since $ 
SU(N) \times U(1)= U(N)\times Z_N$, where $Z_N$ is the discrete permutation group of $N$ 
elements, the orbit of the maximally entangled states can be written as $ {\cal O}_{\rm 
max}=SU(N)/Z_N$. This structure follows also from the fact that the entire orbit may be written 
as ${\cal O}_{\rm max}= (U\otimes {\mathbb I})|\Psi\rangle$, where $|\Psi\rangle$ is an 
arbitrary maximally entangled state, and U is an arbitrary unitary matrix determined up to an 
overall phase \cite{VW00}. 

\subsection{ Special cases: $N=2,3$ and $4$}

The set of all possible Schmidt vectors $\Lambda$ form the $N-1$ dimensional
simplex ${\cal S}_N$. Its corners represent $N$ mutually orthogonal
separable
states, while its center denotes the maximally entangled state
$|\psi_*\rangle=(\sum_{k=1}^N |kk\rangle)/\sqrt{N}$. Any permutation of the
Schmidt coefficients may be obtained by a local transformation of the pure
state. Therefore it is sufficient to consider the orbits generated by
Schmidt
vectors belonging to a certain asymmetric part ${\tilde{\cal S}}_N$ of the
simplex, so called {\sl Weyl chamber}. Any ordering of the Schmidt
coefficients
corresponds to choosing one chamber out of $N!$, in which the simplex ${\cal
S}_N$ can be decomposed.


The Schmidt simplex and exemplary Weyl chamber for $N=2,3$ and $4$ are
presented in Fig.1. (Note that the simplex of diagonal density matrices of size
$N$, obtained from  $N \times N$ pure  states by partial tracing,
has the same geometry).  
The numbers by each part of the boundary of ${\tilde{\cal S}}_N$ denote the
dimensions of the
local orbits, which are listed in Table 1. In the simplest case $N=2$ the
simplex reduces to the
interval $[0,1]$, while its asymmetric part ${\tilde{\cal S}}_2$ equals to
$[0,1/2]$. The edge
$0$ generates the four dimensional orbit of separable states, $ {\mathbb
C}P^{1} \times {\mathbb
 C}P^{1}$, and the point $1/2$ leads to the $3$--D manifold of maximally
entangled states $ {\cal
O}_{\rm max}=SU(2) / Z_2 \approx SO(3) \approx{\mathbb R}P^3$. This
structure was pointed out by
Vollbrecht and Werner \cite{VW00}, and the above singular foliation of
${\mathbb C}P^{3}$ was
discussed in \cite{KZ01,BH01,BBZ01,MD01}. In the case of any point
inside the simplex
(\ref{orbsep}) gives the following topology of the  generic $ 2 \times 2$ 
local orbit
 \begin{equation}
 {\cal O}_{g}=\frac{U(2)}{U(1)^{2}}\times \frac{U(2)}{U(1)}=S^{2}
\times{\mathbb R}P^{3}, 
\end{equation}
in agreement with recent results of Mosseri and Dandoloff \cite{MD01}.

\section{Algebraic determination of orbit dimension}
\label{sec:Gram}

\subsection{General case: $N \times N$ mixed states}

The reasoning presented in the previous section hinges on the Schmidt
decomposition of the density matrix for a pure state. As such it cannot be
extended to mixed states. For this reason we present an alternative method
introduced in \cite{KZ01}, which can be, in principle, applied also in the
latter situation. It is based on purely algebraic reasoning, and, as such,
gives only local information, i.e. only about the dimensions of the
manifolds
of interconvertible states and not about their topology.

Although the group of local unitary transformations is ${\mathcal L}=U(N)\otimes U(N)$, it is 
obvious that since its elements act on an arbitrary density matrix $\rho\in{\mathbb 
C}^N\otimes{\mathbb C}^N$ by conjugations, $\rho\mapsto U\rho U^\dagger$, we can take in fact 
${\mathcal L}=SU(N)\otimes SU(N)$ instead.  Let ${\mathbb R}^{2(N^2-1)}\ni {\mathbf s}\mapsto 
U({\mathbf s})\in SU(N)\otimes SU(N)$ be some parameterization of the group ${\mathcal L}$ such 
that $U(0)=I$ (i.e.\ ${\mathbf s}=\left(s_1,s_2,\ldots,s_{2N^2-2}\right)$ are the coordinates in 
$SU(N)$ with the origin at the unit matrix). The tangent space to the local orbit through $\rho$ 
(i.e.\ to the space of the states interconvertible with $\rho$) at this point is spanned by the 
vectors: 
\begin{equation}
\rho_k:=\frac{\partial}{\partial s_k}U({\mathbf s})\rho U^\dagger({\mathbf
s})
\left.\right|_{{\mathbf s}=0}.
\label{tvs}
\end{equation}
The dimension of the tangent space, hence of the manifold itself, equals the
number of linearly independent vectors $\rho_k$.

From the unitarity of $U({\mathbf s})$ it follows:
\begin{equation}
\rho_k=\left[\left(\frac{\partial U}{\partial s_k}\right)_{{\mathbf
s}=0},\rho\right]=\left[l_k,\rho\right]=\rho_k^\dagger. \label{tvs1}
\end{equation}

The number of independent $\rho_k$ equals the rank of the $2(N^2-1)\times
2(N^2-1)$ Gram matrix (the unimportant factor of $1/4$ is introduced for
further convenience)
\begin{equation}
G_{mn}:=\frac{1}{4}{\mathrm Tr}\rho_m\rho_n,
\label{gram}
\end{equation}
which, upon using (\ref{tvs1}), can be cast into:
\begin{equation}\label{gram1}
G_{mn}  = \frac{1}{2}Tr\left( {l_m \rho l_n \rho } \right) -
\frac{1}{4}Tr\left( {\rho ^2 \left\{ {l_n l_m  + l_m l_n } \right\}}
\right).
\end{equation}

Choosing the standard parameterization of $SU(N)$ in the vicinity of the
identity we obtain
\begin{equation}
l_k:=\left(\frac{\partial U}{\partial s_k}\right)_{{\mathbf
s}=0}=\left\{
\begin{array}{cc}
ie_k\otimes I, & k=1,\ldots,N^2-1 \\
iI\otimes e_k, & k=N^2,\ldots,2N^2-2,
\end{array}
\right.
\label{lk}
\end{equation}
where $e_k=-e_k^\dagger$ are generators of the Lie algebra ${\mathfrak
su}(N)$.
They obey the commutation relations
\begin{equation}
[e_j,e_k]=c_{jkl}e_l,
\label{comm}
\end{equation}
where $c_{jkl}$ denote the structure constants and we use the summation
convention.
We normalize $e_k$ to fulfill
\begin{equation}
{\mathrm Tr\,}e_je_k=-2\delta_{jk}.
\label{kill}
\end{equation}

An arbitrary hermitian matrix $\rho$ acting in ${\mathbb C}^N\otimes{\mathbb
C}^N$ can be decomposed with the help of ${\mathfrak su}(N)$ generators
\begin{equation}
\rho:=\frac{1}{N^2}I+ia_k(e_k\otimes I)+ib_l(I\otimes e_l)+C_{mn}(e_m\otimes
e_n).
\label{dec}
\end{equation}
From (\ref{tvs1}), (\ref{lk}), (\ref{comm}), (\ref{kill}),  and (\ref{dec}) the Gram matrix 
(\ref{gram}) is calculated as 
\begin{equation}
G=\left[
\begin{array}{lr}
A & B \\
B^T & D
\end{array}
\right],
\label{C}
\end{equation}
where the $(N^2-1)\times (N^2-1)$ matrices $A,B$, and $D$ read
\begin{eqnarray}
A_{mn}&=&c_{mjk}c_{nlk}(Na_ja_l+2C_{jr}C_{lr})/2, \nonumber \\
B_{mn}&=&c_{mjk}c_{nlr}C_{kl}C_{jr}, \nonumber \\
D_{mn}&=&c_{mjk}c_{nlk}(Nb_jb_l+2C_{rj}C_{rl})/2.
\end{eqnarray}

\subsection{Special case:  $N \times N$ pure states}

Using the above outlined procedure we can recover the results for pure
states
obtained in Section I. For a pure state $\rho=\left| \psi  \right\rangle
\left\langle \psi  \right|$ Eq.~(\ref{gram1}) reduces to
\begin{equation}\label{grampure}
G_{mn}  = \left\langle\psi \right|l_m\left|\psi\right\rangle
\left\langle\psi\right|l_n\left|\psi\right\rangle- \frac{1}{2}\left\langle
\psi
\right|l_m l_n + l_n l_m \left| \psi\right\rangle\left\langle
\psi|\psi\right\rangle.
\end{equation}
We choose the following explicit form of the generators $e_k$ expressed in
the
standard basis $\{\left| 1 \right\rangle ,\left| 2 \right\rangle ,\cdots ,
\left| N \right\rangle\}$ of ${\mathbb C}^N$
\begin{gather}\label{ecart}
e_k = - i\sqrt {\frac{2} {{k(k + 1)}}} \left( {k\left| k+1 \right\rangle
\left\langle k+1 \right| - \sum\limits_{l = 1}^k {\left| l \right\rangle }
\left\langle l \right|} \right), \quad k=1,2,\ldots,N-1,
\\
e_{mn}^{(1)}= i(\left| n \right\rangle \left\langle m \right| + \left| m
\right\rangle \left\langle n \right|), \quad 1\leq m<n\leq N,
\\
 e_{mn}^{(2)}=
\left| n \right\rangle \left\langle m \right| - \left| m \right\rangle
\left\langle n \right|, \quad 1\leq m<n\leq N.
\end{gather}
We reorder the non-diagonal generators $e_{mn}^{(1)}$ and $e_{mn}^{(2)}$ by
changing two indices $\{mn\}$ into a single one $k$ according to
$k=N-1+(m-1)N-
m(m+1)/2+n$ in the case of $e_{mn}^{(1)}$ and $k=N-1+N(N-1)/2-m(m+1)/2+n$ in
the case of $e_{mn}^{(2)}$, so that $\{e_k\}$, $k=1,2,\ldots,N^2-1$ is the
desired complete set of generators.

It proves to be more convenient to use not $l_k$ themselves, but the
following
linear combinations of them:
\begin{eqnarray}\label{lm}
L_k&:=&i(e_k\otimes I+I\otimes e_k)/2, \quad 1\leq k\leq N^2-1, \\
L_k&:=&i(e_k\otimes I-I\otimes e_k)/2, \quad N^2\leq k\leq 2N^2-2,
\end{eqnarray}
what amounts to a mere change of basis in the Lie algebra and, obviously,
does
not influence the rank of $G$.

After rather straightforward but lengthy calculation we find $G$ in the form (\ref{C}) with 
$B=0$ and bloc-diagonal matrices $A$ and $D$ 
\begin{eqnarray}\label{AD}
  A=\left[
  \begin{array}{lr}
  A^{(1)} & 0 \\
  0 & A^{(2)}
  \end{array}
  \right], \quad
   D=\left[
  \begin{array}{lr}
  D^{(1)} & 0 \\
  0 & D^{(2)}
  \end{array}
  \right].
\end{eqnarray}
The blocks $A^{(2)}$ and $D^{(2)}$ are diagonal $(N^2-N)\times(N^2-N)$
matrices
with the diagonal entries
\begin{eqnarray}
A^{(2)}_{kk}&=&(\sqrt{\lambda_m}+\sqrt{\lambda_n})^2\left(\sum_{j=1}^N \lambda_j\right), \quad 
1\leq k\leq (N^2-N)/2, \label{G2d1} \\ 
A^{(2)}_{kk}&=&(\sqrt{\lambda_m}-\sqrt{\lambda_n})^2\left(\sum_{j=1}^N \lambda_j\right), \quad 
(N^2-N)/2<k\leq N^2-N, \label{G2d2} \\ 
D^{(2)}_{kk}&=&(\sqrt{\lambda_m}-\sqrt{\lambda_n})^2\left(\sum_{j=1}^N \lambda_j\right), \quad 1 
\leq k\leq(N^2-N)/2, \label{G2d3} \\ 
D^{(2)}_{kk}&=&(\sqrt{\lambda_m}+\sqrt{\lambda_n})^2\left(\sum_{j=1}^N \lambda_j\right), \quad 
(N^2-N)/2<k\leq N^2-N, \label{G2d4}. 
\end{eqnarray}
In each of the above formulas $(m,n)$ is the unique pair of numbers such
that
$0<m<n\leq N$ and fulfilling $(m-1)N-m(m+1)/2+n=k$ for $1\leq k\leq
(N^2-N)/2$
or $(m-1)N-m(m+1)/2+n=k-(N^2-N)/2$ for $(N^2-N)/2<k\leq N^2-N$. Moreover, we
find that of two $(N-1)\times (N-1)$ matrices $A^{(1)}$ and $D^{(1)}$ the
latter equals zero, while the former reads
\begin{eqnarray}\label{Mc}
A^{(1)}_{mn}&=&\frac{{\left( {\sum\nolimits_{k = 1}^m {\lambda _k  -
m\lambda
_{m+1} } } \right)\left( {\sum\nolimits_{k = 1}^N {\lambda _k } } \right) -
\left( {\sum\nolimits_{k = 1}^m {\lambda _k  - m\lambda _{m+1} } }
\right)\left( {\sum\nolimits_{k = 1}^n {\lambda _k  - n\lambda _{n+1} } }
\right)}} {{\sqrt {m(m + 1)n(n + 1)} }}=A^{(1)}_{nm}, \quad m<n,
\\
A^{(1)}_{nn}&=& \frac{{\left( {\sum\nolimits_{k = 1}^n {\lambda _k  + n^2
\lambda _{n + 1} } } \right)\left( {\sum\nolimits_{k = 1}^N {\lambda _k } }
\right) - \left( {\sum\nolimits_{k = 1}^n {\lambda _k  - n\lambda _{n +
1} } }
\right)^2 }} {{n(n + 1)}}\,.
\end{eqnarray}
In this way we found that the entire matrix $G$ has at least $N-1$ vanishing
eigenvalues (due to $D^{(1)}=0$), $N^2-N$ doubly degenerate eigenvalues
$(\lambda_i\pm\lambda_j)^2$ (the eigenvalues of $A^{(2)}$ and $D^{(2)}$) and
the $N-1$ eigenvalues of $A^{(1)}$.

Although, at first sight, $A^{(1)}$ looks quite complicated, it is
relatively
easy to calculate the traces of its powers $\text{Tr}(A^{(1)})^k$,
$k=1,2,\ldots,N-1$ and, consequently, its characteristic polynomial
\begin{equation}
P\left(
\lambda\right):=\text{det}(A^{(1)}-\lambda)=\sum_{k=1}^{N}\left(-1\right)^{k
+1}k
p_{k}\lambda^{N-k}. \label{P}
\end{equation}
Here $p_1=\tau_1$, $p_2=\tau_2$, and
$p_k=\tau_k\left(\sum_{j=1}^N\lambda_j\right)^{k-2}$ where $\tau_k$ are the
coefficients of
\begin{equation}
Q\left(\lambda\right):=\prod_{i=1}^{N}\left(\lambda-\lambda_{i}\right)
=\sum_{k=1}^{N}\left(-1\right)^{k}\tau_{k}\lambda^{N-k} \label{Q},
\end{equation}
i.e. the elementary symmetric polynomials in $\lambda_1,\lambda_2,\ldots,
\lambda_N$ of the order $k$. Observe that due to the normalization $
\sum_{k=1}^{N} \lambda_k=1$ we can substitute $\tau_k$ for $p_k$ in
(\ref{P})
and, consequently,
\begin{equation}\label{Pq}
  P\left( \lambda\right)=\sum_{k=1}^{N}\left(-1\right)^{k+1}k
p_{k}\lambda^{N-k}=\lambda Q^{\prime}\left(\lambda\right)
-NQ\left(\lambda\right),
\end{equation}
where $Q^{\prime}\left(\lambda\right):={dQ\left(\lambda\right)}/{d\lambda}$.
It
follows immediately that the multiplicity of the root $\lambda=0$ in $P$
equals
the multiplicity of $\lambda=0$ in $Q$ (i.e.\ the number of Schmidt
coefficients equal to $0$). Indeed, if $Q(\lambda)=\lambda^kQ_1(\lambda)$
and
$Q_1(0)\ne0$ then
$Q^{\prime}(\lambda)=k\lambda^{k-1}Q_1(\lambda)+\lambda^kQ_1^{\prime}
(\lambda)$
and $P(\lambda)=\lambda^k\left[(k-N)Q_1(\lambda)+\lambda
Q_1^{\prime}(\lambda)\right] =\lambda^kP_1(\lambda)$, where
$P_1(0)=(k-N)Q_1(0)\ne 0$ since $k\leq N$.

Now we are ready to calculate the rank of $G$. There are
\begin{enumerate}
\item $N-1$ vanishing eigenvalues of $D^{(1)}$,
\item $m_0$ vanishing eigenvalues of $A^{(1)}$,
\item for each $m_n$-degenerate Schmidt coefficient $m_n(m_n-1)$ vanishing
eigenvalues of $A^{(2)}$ of the form (\ref{G2d2}) and of the form
(\ref{G2d3})
of $D^{(2)}$,
\item $2m_0(m_0-1)$ vanishing eigenvalues of $A^{(2)}$ and $D^{(2)}$ of the
forms (\ref{G2d1}--\ref{G2d4}).
\end{enumerate}
hence the co-rank (the number of zero eigenvalues of $G$) equals
$(N-1)+m_0+\sum_{n=1}^K(m_n^2-m_n)+2(m_0^2-m_0)=2m_0^2+\sum_{n=1}^Km_n^2$-1,
where we used $m_0+\sum_{n=1}^Km_n=N$. Consequently, taking in account that
$G$ is an $2(N^2-1)\times 2(N^2-1)$ matrix, its rank equal to
the dimension of the orbit is given by (\ref{dimorb}).

As mentioned at the beginning of the section, the above analysis can be, in
principle, extended to mixed states. To show this let's consider (admittedly
rather trivial) example of the generalized Werner states
\begin{equation}\label{werner}
  \rho=\frac{1-\alpha}{N}I+\alpha|\psi\rangle\langle\psi|,
\end{equation}
where the pure state $|\psi\rangle$ is characterized by the Schmidt numbers
$(0\leq\lambda_1\leq\lambda_2\leq\cdots \leq\lambda_N)$. It is obvious that
the
$\rho_k$ of Eq.~(\ref{tvs1}) are, up to the scaling factor $\alpha$ the same
as
for the pure state $|\psi\rangle$. Consequently, the dimension of the orbit
through $\rho$ is determined by the Schmidt coefficients of $|\psi\rangle$
exactly in the same way as previously.

\section{Coefficients of the characteristic polynomials as
  entanglement measures}

There exist several non equivalent ways to quantify quantum entanglement
\cite{HHH00,VP00,DHR01}. Following Vedral and Plenio \cite{VP98} we assume that
any entanglement measure

i) equals to zero for any separable state,

ii) is invariant with respect to local unitary operations,

iii)  cannot increase under operations involving local
measurements  and classical communication.

For pure states, $\rho=|\psi\rangle\langle\psi|$, these requirements are
fulfilled by the Shannon entropy of the Schmidt vector, (in other words von
Neumann entropy of the partially reduced density matrix),
$E_1(|\psi\rangle)=-\sum_{k=1}^N \lambda_k \ln \lambda_k $, simply called
{\sl
entropy of entanglement}, as well as the generalized Renyi entropies,
$E_{\alpha}(|\psi\rangle) = {\mathrm ln}(\sum_{k=1}^N \lambda_k^{\alpha}
)/(1-\alpha)$ \cite{Vi00,ZB01}.

Consider now the coefficients $\tau_k$ of the characteristic polynomial
(\ref{P}) of the nontrivial block $A^{(1)}$ of the Gram matrix
(\ref{grampure})
for a pure state of a $N \times N$ bipartite system. As derived above they
are
given by the elementary symmetric polynomials in
$\lambda_1,\lambda_2,\ldots,
\lambda_N$ of the order $k$
\begin{eqnarray}
\label{tautau}
\tau_1 &=&  \sum_{k=1}^N\lambda_k=1, \nonumber \\
\tau_2 &=&  \sum_{k=1}^N \sum_{l=k+1}^N \lambda_k\lambda_l, \nonumber \\
\tau_3 &=&  \sum_{k=1}^N \sum_{l=k+1}^N \sum_{m=l+1}^N
\lambda_k\lambda_l\lambda_m,  \nonumber \\
... & & ...\nonumber \\
\tau_N &=&  \prod_{k=1}^N\lambda_k.
\end{eqnarray}
Due to the definition of the Gram matrix the coefficients $\tau_k$,
$k=2,\dots,N$ are invariant with respect to local unitary transformations
and are equal to zero if and only if the state is separable.

As shown recently by Nielsen \cite{Ni99} any pure state $|\psi\rangle$ may
be
transformed locally into a given state $|\phi\rangle$, if and only if the
corresponding vectors of the Schmidt coefficients satisfy the following
majorization relation ${\vec \lambda}_{\psi} \prec {\vec \lambda}_{\phi}$.
Any
entanglement measure cannot increase under such an operation. This condition
is
fulfilled by the coefficients $\tau_k$, since the elementary symmetric
polynomials are known to be {\sl Schur--concave} functions \cite{MO79}, for
which ${\vec \lambda} \prec {\vec \mu}$ induces $\tau({\vec \lambda}) \ge
\tau({\vec \mu})$. Thus the quantities (\ref{tautau}) posses the property of
{\sl entanglement monotones}, and their set consisting of $N-1$ independent
elements,
 $\{\tau_2,\dots\tau_N \}$, provides the complete characterization of the
pure
states entanglement \cite{Vi00}. Beside the simplest case of $N=2$, (for
which all measures of
the entanglement generate the same order in the set of pure states
\cite{ZB01}), the
coefficients $\tau_k$ are not functions of the Renyi entropies and induce
different orders in
the set of pure entangled states.

It might be interesting to analyze how the traces of the Gram matrix,
$t_k:={\mathrm tr}(G^k)$, change during non-unitary local transformations.
Our
numerical experiments performed for mixed states of $2\times 2$ system
suggest
that all traces $t_k$, $k=1,\ldots,6$ do not increase under local
bistochastic
transformations , $\rho\mapsto\rho^{\prime}=\sum_i p_iU_i^A\otimes U_i^B\rho
U_i^{A\dagger}\otimes U_i^{B\dagger}$, with $\sum_i p_i=1$. The question
whether
this property holds also for systems of higher dimensions remains open.


\acknowledgments It is a pleasure to thank I.~Bengtsson, D.~C.~Brody, P.~Heinzner, 
A.~Huckelberry, J.~Kijowski and J.~Rembieli\'nski for fruitful discussions and R. Mosseri for 
helpful correspondence. The work was supported by Polish Komitet Bada\'n Naukowych through 
research Grant No 2 P03B 072 19.


\bigskip    
\begin{figure}
\includegraphics{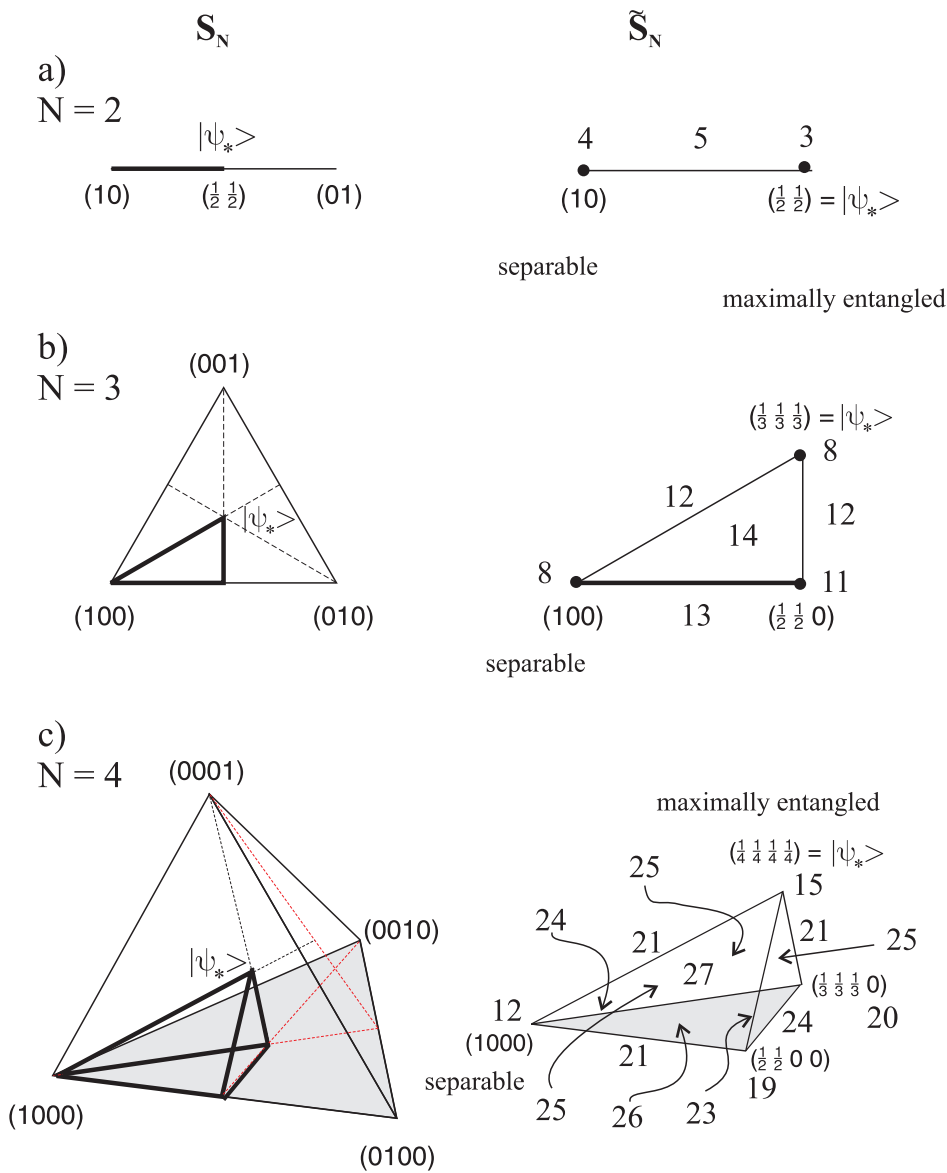}
\caption{Simplex of Schmidt coefficients $S_N$ 
 for pure states of $N\times N$ system with $N=2,3$, and $4$;
 (the same picture may also represent the set of the spectra of density
matrices of size $N$ obtained from pure states by partial tracing). Right hand side shows an 
asymmetric part of $S_N$ - the Weyl chamber ${\tilde S}_N$, while the numbers denote the 
dimensionality of local orbits generated by each point.} 
\end{figure}      
\renewcommand{\arraystretch}{1.7}
{\large 
\begin{tabular}[c]{|c|c|c|c|cccl|c|}
 \hline 
 \raisebox{-2ex}{$N$} & \parbox {4cm}{\centering Schmidt  } & \raisebox{-2ex}{$D_s$} & \parbox {4cm}{\centering Part of the  } &
\multicolumn{4}{|c|}{ Topological Structure }& \raisebox{-2ex}{$D_o$} \\ 
 & coefficients && asymmetric simplex & {\it base } & &  {\it fibre}  && \\ \hline\hline
    & $(a,b)$ & $1$ & line  & $\frac{U(2)}{[U(1)]^{2}}$&$ \times$&$ \frac{U(2)}{U(1)}$&$=S^{2} 
    \times {\mathbb R} P^{3}$ & $5$\\\cline{2-8}
 $2$ & $(1,0)$ & $0$ & left edge ( $\circ$ )& $\frac{U(2)}{[U(1)]^{2}}$&$ \times$&$ 
\frac{U(2)}{U(1)\times U(1) }     
 $&$={\mathbb C}P^{1}\times {\mathbb C}P^{1}$ & $4$\\\cline{2-8}
  & $(1/2,1/2)$ & $0$ & right edge ( $\star$ ) & $\frac{U(2)}{U(2)} $&$\times$&$ \frac{U(2)}{U(1)}$&$
  =\frac{SU(2)}{Z_{2}}={\mathbb R}P^{3}$ & $3$\\\hline\hline
& $(a,b,c)$ & $2$ & \parbox{4cm}{\centering interior of triangle} & $\frac{U(3)}{[U(1)]^{3}}$&$ 
\times$&$ \frac{U(3)}{U(1)}$  && $14$\\\cline{2-8} & $(a,b,0)$ & $1$ & base & 
$\frac{U(3)}{[U(1)]^{3}} $&$\times$&$ \frac{U(2)}{U(1)\times U(1)}$ &  & $13$\\\cline{2-3} $3$ & 
$(a,b,b)$ & $1$ & 2 upper sides & $\frac{U(3)}{U(1)\times U(2)} $&$\times $&$\frac{U(3)}{U(1)}$& 
& $12$\\\cline{2-8} & $(1/2,1/2,0)$ & $0$ & right corner  & $\frac{U(3)}{U(2)\times U(1)} 
$&$\times$&$ \frac{U(3)}{U(1)\times U(1)}$& & $11$\\\cline{2-3} & $(1,0,0)$ & $0$ &  left corner     
( $\circ$ ) & $\frac{U(3)}{U(1)\times U(2)} $&$\times $&$\frac{U(3)}{U(2)\times 
U(1)}$&$={\mathbb C}P^{2}\times {\mathbb C}P^{2}$ & $8$ 
\\\cline{2-3}
& $(1/3,1/3,1/3)$ & $0$ &  upper corner ( $\star$ )& $\frac{U(3)}{U(3)}$&$\times$&$ 
\frac{U(3)}{U(1)} $&$= \frac{SU(3)}{Z_{3}}$ & $8$\\ \hline\hline                                                       
& $(a,b,c,d)$ & $3$ & \parbox{4cm}{\centering interior of \raisebox{.4ex}{tetrahedron }} & 
$\frac{U(4)}{[U(1)]^{4}}$&$\times $&$ \frac{ U(4)}{U(1)}$ && $27$\\\cline{2-8} 
 & $(a,b,c,0)$ & 
$2$ & base  face & $\frac{U(4)}{[U(1)]^{4}} $&$\times$&$ \frac{U(4)}{[U(1)]^{2}}$ && 
$26$\\\cline{2-3}
  & $(a,a,b,c)$ & $2$ & three upper faces & $\frac{U(4)}{ U(2)\times [U(1)]^{2} 
} $&$\times $&$ \frac{U(4)}{U(1)}$ && $25$\\\cline{2-8} & $(a,a,b,0)$ & $1$ & 2 edges of the 
base &$\frac{U(4)}{ U(2)\times [U(1)]^{2} }$&$ \times $&$ \frac{U(4)}{U(1)\times U(1)}$ && 
$24$\\\cline{2-3} $4$ & $(a,a,b,b)$ & $1$ & edge & $\frac{U(4)}{[U(2)]^{2} } $&$\times $&$ 
\frac{U(4)}{U(1)}$  && $23$\\\cline{2-3} 
 & $(a,a,a,b)$ & $1$ & 2 edges & $\frac{U(4)}{U(3) \times U(1)} $&$\times $&$ \frac{U(4)}{U(1)}$
  &&$21$\\\cline{2-3}
 & $(a,b,0,0)$ & $1$ & lower edge of the base & $\frac{U(4)}{[U(1)]^{2} \times U(2)} $&$\times 
$&$ \frac{U(4)}{U(2)\times U(1)}$  && $21$\\\cline{2-8} & $(1/3,1/3,1/3,0)$ & $0$ & back corner 
&$\frac{U(4)}{U(3) \times U(1)} $&$\times $&$ \frac{U(4)}{U(1)\times U(1)}$  && 
$20$\\\cline{2-3} 
 & $(1/2,1/2,0,0)$ & $0$ & right corner & $\frac{U(4)}{[U(2)]^{2}} $&$\times $&$ 
\frac{U(4)}{U(2)\times U(1)}$  && $19$\\\cline{2-3}
 & $(1/4,1/4,1/4,1/4)$ & $0$ & upper corner ( $\star$ )  & $ \frac{U(4)}{U(4)}$&$ \times  $&$
\frac{U(4)}{U(1)} $&$ =\frac{SU(4)}{Z_{4}}$ & $15$\\\cline{2-3} & $(1,0,0,0)$ & $0$ & left 
corner ( $\circ$ ) & $\frac{U(4)}{U(1) \times U(3)} $&$\times  $&$ \frac{U(4)}{U(3)\times 
U(1)}$&$={\mathbb C}P^{3}\times {\mathbb C}P^{3}$ & $12$\\\hline 
\end{tabular}

\bigskip

Table 1. \ Topological structure of local orbits of the $N\times N$ \ pure states generated by 
one Weyl chamber of the simplex  of the Schmidt coefficients, $D_s$ is the dimension of the 
subspace, while $D_o$ represents the dimension of the orbit, ( $\circ$ ) denotes separable 
states, while ( $\star$ ) denotes maximally entangled states.  }                                          
\end{document}